\documentclass[aps,prb,twocolumn,showpacs,superscriptaddress]{revtex4}

\usepackage{graphicx}
\usepackage{bm}

\begin{document}

\preprint{}

\title{Emergent Phases of Nodeless and Nodal Superconductivity Separated by Antiferromagnetic Order in Iron-based Superconductor (Ca$_4$Al$_2$O$_{6}$)Fe$_2$(As$_{1-x}$P$_x$)$_2$:\\ $^{75}$As- and $^{31}$P-NMR Studies}

\author{H. Kinouchi}\email[]{e-mail  address: kinouchi@nmr.mp.es.osaka-u.ac.jp}
\author{H. Mukuda}\email[]{e-mail  address: mukuda@mp.es.osaka-u.ac.jp}
\author{Y. Kitaoka}
\affiliation{Graduate School of Engineering Science, Osaka University, Toyonaka, Osaka 560-8531, Japan}

\author{P. M. Shirage}
\author{H. Fujihisa}
\author{Y. Gotoh}
\author{H. Eisaki}
\author{A. Iyo}
\affiliation{National Institute of Advanced Industrial Science and Technology (AIST), Umezono, Tsukuba 305-8568, Japan}

\date{\today}

\begin{abstract}
We report $^{31}$P- and $^{75}$As-NMR studies on (Ca$_4$Al$_2$O$_{6}$)Fe$_2$(As$_{1-x}$P$_x$)$_2$ with an isovalent substitution of P for As. We present the novel evolution of emergent phases that the nodeless superconductivity (SC) in  
0$\le x \le$0.4 and the nodal one around $x$=1 are intimately separated by the onset of a commensurate stripe-type antiferromagnetic (AFM) order in  0.5$\le x \le$ 0.95, as  an isovalent substitution of P for As decreases a pnictogen height $h_{Pn}$ measured from the Fe plane. It is demonstrated that 
the AFM order takes place under a condition of 1.32\AA $\le h_{Pn} \le$1.42\AA , which is also the case for other Fe-pnictides with the Fe$^{2+}$ state in (Fe$Pn$)$^{-}$ layers. This novel phase evolution with the variation in $h_{Pn}$ points to the importance of electron correlation for the emergence of SC as well as AFM order. 
\end{abstract}


\maketitle


High-transition-temperature ($T_{\rm c}$) superconductivity (SC) in iron (Fe)-pnictides ($Pn$)\cite{Kamihara} emerges when a stripe-type antiferromagnetic (AFM) order accompanied by a structural phase transition is suppressed by either a carrier doping, an application of pressure, etc.
The parent compounds are AFM semimetal characterized by an average iron valence Fe$^{2+}$ in (Fe$Pn$)$^-$ layer separated by various block layers. These compounds are $Ln$FeAsO ("1111") ($Ln$=rare earth)\cite{Kamihara,Cruz,Qui}, $Ae$Fe$_2$As$_2$ ("122") ($Ae$=Alkaline earth)\cite{RotterM,Huang}, and ($Ae_4M_2$O$_6$)Fe$_2$As$_2$ ("42622")\cite{Sato,Yamamoto,Munevar}. By contrast, the onset of AFM order has not yet been reported in Fe-phosphorus parent compounds with the Fe$^{2+}$ in (FeP)$^-$ layer.  Although the parent compound NaFeAs ("111") exhibits an AFM order~\cite{Chen,Ma,Kitagawa}, whereas  a fully gapped or nodeless SC emerges in  LiFeAs without any carrier doping~\cite{Wang,Tapp,Kim,Hashimoto}.  Meanwhile, the isovalent substitution of P for As in BaFe$_2$(As$_{1-x}$P$_x$)$_2$ (hereafter denoted as Ba122(As, P)) brings about the SC with nodal gap~\cite{Kasahara,NakaiPRB,Yamashita,Zhang}. 
Thus, the compounds with the Fe$^{2+}$ state in (Fe$Pn$)$^{-}$ layers undertake an intimate evolution into either the nodeless SC in LiFeAs and (Ca$_4$Al$_2$O$_{6}$)Fe$_2$As$_2$\cite{Kinouchi} or the nodal SC in Ba122(As,P) without any change in the valence condition of the Fe$^{2+}$ in the (Fe$Pn$)$^-$ layer. To gain further insight into a novel phase evolution when the Fe$^{2+}$ state is kept in the (Fe$Pn$)$^-$ layer, we have dealt with (Ca$_4$Al$_2$O$_{6}$)Fe$_2$(As$_{1-x}$P$_x$)$_2$ (hereafter denoted as Al-42622(As,P)) in which the Fe$^{2+}$ state is expected  irrespective of the P-substitution for As in the (Fe$Pn$)$^-$ layer separated by a thick perovskite-type block\cite{Shirage,Kinouchi}. Here, note that a highly two-dimensional electronic structure in these compounds is in contrast with the three-dimensional one observed in $Ae$122(As,P)($Ae$=Ba,Sr)~\cite{Kasahara,Zhang,Dulguun}.

In this Letter, we report on a novel phase diagram for  Al-42622(As,P) with the isovalent substitution of P for As and hence without any carrier doping.  $^{31}$P- and $^{75}$As-NMR studies have revealed that a commensurate AFM order taking place in $0.5 \le x \le 0.95$ intervenes between a nodeless SC in $0 \le x \le 0.4$ and a nodal one around $x$=1.  We highlight that as the substitution of P for As decreases a pnictogen height $h_{Pn}$ measured from the Fe plane, the nodeless SC state evolves into an AFM-order state and subsequently into a nodal SC state, while keeping the Fe$^{2+}$ state due to the isovalent substitution of P for As. We remark that this finding points to the importance of the electron correlation effect for the emergence of SC as well as  AFM order in Fe-pnictides in general.

\begin{figure}[thbp]
\centering
\includegraphics[width=8cm]{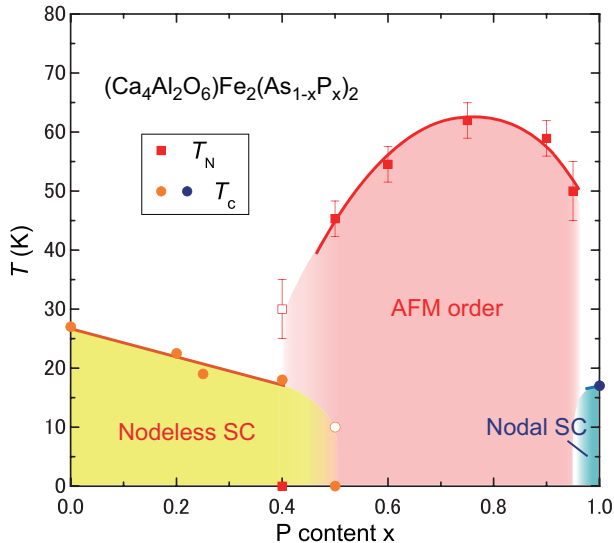}
\caption[]{(Color online) (a) Phase diagram against $x$ in (Ca$_4$Al$_2$O$_{6}$)Fe$_2$(As$_{1-x}$P$_x$)$_2$. 
The commensurate AFM order occurs in 0.5$\le x \le$ 0.95, which intervenes between the nodeless SC in 0$\le x \le$ 0.4 and the nodal SC at $x$=1. Each empty symbol near the phase boundary between the nodeless SC phase and the AFM ordered phase means that each sample contains a tiny fraction of minority domain exhibiting either AFM order or nodeless SC due to a possible spatial distribution of As/P content. 
}
\label{Phase}
\end{figure}

Polycrystalline samples of (Ca$_4$Al$_2$O$_{6-y}$)Fe$_2$(As$_{1-x}$P$_x$)$_2$ with a nominal content of $0\leq x\leq 1$ were synthesized by the solid-state reaction method using the high-pressure synthesis technique described elsewhere~\cite{Shirage,Shirage_AsP}. Due to the oxidization of the starting materials, a nominal value of $y$ in the prepared samples may be empirically nearly zero, even though $y\sim 0.2$.  
Powder X-ray diffraction measurements indicate that these samples are almost entirely composed of a single phase, and the lattice parameters such as the lengths along $a$-axis and  $c$-axis, and $h_{Pn}$ at room temperature decrease monotonously with increasing $x$, ensuring a homogeneous chemical substitution of P for As. 
Here, $h_{Pn}$s for $x$=0 and 1 were obtained from Rietveld analyses, and those in the intermediate $x$ region were tentatively deduced from a linear interpolation from $x$=0 to 1~\cite{Supp.}. 
Bulk $T_{\rm c}$s for $0 \le x \le 0.4$ and $x$=1.0 were determined from an onset of SC diamagnetism in the susceptibility measurement, whereas no SC transition was identified in $0.5 \le x \le 0.95$ (see Fig.~\ref{Phase})~\cite{Shirage_AsP}. 
$^{31}$P-NMR($I=1/2$) measurements have been performed on coarse powder samples.
The nuclear spin-lattice relaxation rate ($1/T_1$) of $^{31}$P-NMR was obtained by fitting a recovery curve of $^{31}$P nuclear magnetization to a single exponential function $m(t)\equiv (M_0-M(t))/M_0=\exp \left(-t/T_1\right)$. 
Here, $M_0$ and $M(t)$ are the respective nuclear magnetizations for a thermal equilibrium condition and at time $t$ after a saturation pulse. 

\begin{figure}[hbp]
\centering
\includegraphics[width=8cm]{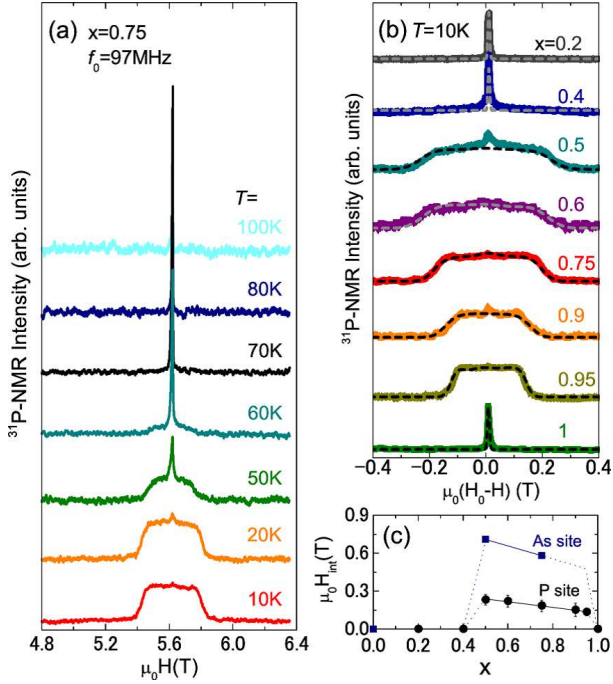}
\caption[]{(Color online) 
(a) $T$ dependence of $^{31}$P-NMR spectrum at $x$=0.75, pointing to the onset of AFM order below 60\,K. (b) $x$ dependences of $^{31}$P-NMR spectrum at 10\,K along with a simulation(broken line) and (c) respective internal fields $^{31}H_{\rm int}$ and $^{75}H_{\rm int}$ at $^{31}$P and $^{75}$As. 
}
\label{NMRspectra}
\end{figure}

Figure~\ref{NMRspectra}(a) shows temperature ($T$) dependence of the $^{31}$P-NMR spectrum at $x$=0.75 at a fixed frequency $f_0$=97 MHz. At  temperatures higher than 70 K, the $^{31}$P-NMR spectrum is composed of a single sharp peak. 
The spectrum becomes significantly broader upon cooling below 60 K and finally exhibits a rectangular-like spectral shape at $\sim$10 K. This characteristic NMR shape is a powder pattern  expected for the case where the $^{31}$P nucleus experiences a uniform off-diagonal internal hyperfine field $H_{\rm int}$ associated with a commensurate stripe-type AFM order of Fe-$3d$ electron spins \cite{Kitagawa1}. The $^{31}H_{\rm int}\simeq$0.19 T at $^{31}$P is estimated at 10 K for $x$=0.75. 
The $^{75}H_{\rm int}\simeq$0.58 T at $^{75}$As is evaluated from $^{75}$As-NMR spectrum (not shown). Here, note that $^{75}H_{\rm int}$ is larger than $^{31}H_{\rm int}$ because of the hyperfine-coupling constant $^{75}A_{\rm hf}$ being larger than $^{31}A_{\rm hf}$. Using the relationship of $^{75}H_{\rm int}$=$^{75}A_{\rm hf}M_{\rm AFM}$, a Fe-AFM moment $M_{\rm AFM}\sim$0.23$\mu_B$ is estimated assuming $^{75}A_{\rm hf}\sim$ 2.5 T/$\mu_B$ which is cited from previous reports~\cite{Kitagawa1,Graf}. In order to present a systematic evolution of the low-$T$ phase as a function of P-substitution, 
the respective figures~\ref{NMRspectra}(b) and ~\ref{NMRspectra}(c) indicate the $x$-dependences of $^{31}$P-NMR spectrum and the internal fields of $^{31}H_{\rm int}$ and $^{75}H_{\rm int}$ at 10 K. 
The $M_{\rm AFM}$ increases from 0.16$\mu_B$ at $x$=0.95 to 0.28$\mu_B$ at $x$=0.5, and becomes zero at $0\le x \le 0.4$ and $x$=1. Note that the $M_{\rm AFM}$s in these compounds were smaller than in another parent 42622 compound [Sr$_4$(MgTi)O$_{6}$]Fe$_2$As$_2$\cite{Yamamoto},  but larger than in (Sr$_4$Sc$_2$O$_6$)Fe$_2$As$_2$~\cite{Munevar}. 

\begin{figure}[thbp]
\centering
\includegraphics[width=7cm]{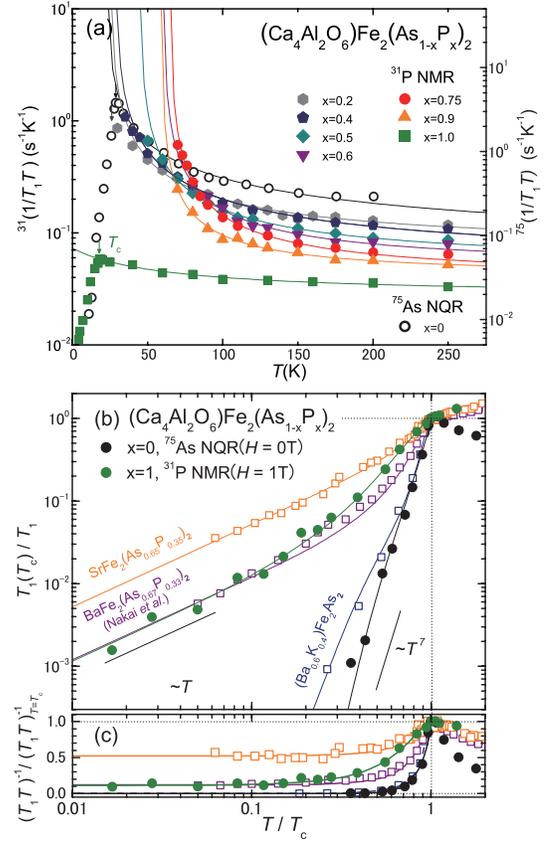}
\caption[]{(Color online) 
(a) $T$ dependence of $^{31}$P-NMR $^{31}(1/T_1T)$ in 0.2$\le x \le$1 along with $^{75}$As-NQR $^{75}(1/T_1T)$ at $x$=0 at the normal state. Note that the $^{31}(1/T_1T)$s for $0.5 \le x \le 0.9$ diverge toward $T_N$, below which the $^{31}$P-NMR spectrum exhibits a rectangular-like shape in association with the onset of an AFM order (see Fig. 1(b)). (b) Plots of $^{31}$NMR - $T_1(T_c)/T_1$ and (c) $(T_1T)^{-1}/(T_1T)^{-1}_{T_c}$ versus $T/T_c$ in the SC state at $x$=1 along with that at $x$=0~\cite{Kinouchi}. The $T$-linear dependence in $1/T_1$ well below $T_c$ at $x$=1 indicates the presence of the residual DOS at $E_F$ in association with the nodal SC in contrast with the nodeless SC at $x$=~0~\cite{Kinouchi}. 
}
\label{1/T1TvsT}
\end{figure}


Figure \ref{1/T1TvsT}(a) shows the $T$ dependence of $^{31}(1/T_1T)$ of $^{31}$P-NMR for $0.2 \le x \le 1$ along with that of $^{75}(1/T_1T)$ of $^{75}$As-NQR for $x$=0. In general, $1/T_1T$ is proportional to $\sum_{\bm q} |A_{\bm q}|^2 \chi''({\bm q},\omega_0)/\omega_0$, where $A_{\bm q}$ is a wave-vector (${\bm q}$)-dependent hyperfine-coupling constant, $\chi({\bm q},\omega)$ a dynamical spin susceptibility, and $\omega_0$ an NMR frequency.  When a system undergoes an AFM order with a wave vector ${\bm Q}$, staggered susceptibility $\chi_{\rm Q}(T)$, following a Curie-Weiss law, diverges towards $T_N$. As a result, the measurement of $1/T_1T$ enables the determination of a N\'eel temperature $T_N$. Actually, as shown by the solid lines in Fig.~\ref{1/T1TvsT}(a), the $1/T_1T$s for $0.2 \le x \le 1$ can be fitted by assuming $1/T_{1}T \sim a/(T-\theta)+b$~\cite{Moriya}. Remarkably, the $1/T_1T$s for 0.4$\le x \le$0.95 diverge toward $T=\theta$, that is identified as $T_N$, because the rectangular-like NMR spectral shape points to the onset of AFM order below $T\sim\theta$, as seen in Fig. \ref{NMRspectra}. In Fig.~\ref{Phase}, the estimated $T_{\rm N}$ is plotted as the function of $x$.

Fig.~\ref{1/T1TvsT}(b) presents the $T$ dependence of $T_1(T_c)/T_1$ of $^{31}$P-NMR in the SC state at $x$=1, which  decreases markedly without any trace of a coherence peak below $T_{\rm c}$, followed by a $T$-linear dependence well below $T_c$. This $T_1T$=const.~behavior at low-$T$ probes the presence of residual density of states (RDOS) at the Fermi level at an external field $H$=1 T, pointing to the SC realized with nodal gap. This nodal SC at $x$=1 is in contrast with the nodeless SC at $x$=0. For $x$=0, the $1/T_1T$ decreases steeply without any coherence peak down to zero toward $T\rightarrow$0 K as presented in Fig. \ref{1/T1TvsT}(c)~\cite{Kinouchi}. 

Generally, the RDOS at $E_F$ is induced for the nodal SC state by either applying $H$ or the presence of some impurity scattering.  Since $1/T_1T$ is related to the square of RDOS at $E_F$ ($N_{\rm res}^2$), the ratio of RDOS ($N_{\rm res}/N_0$) to a normal-state DOS ($N_0$) is given by $\sqrt{(T_1T)^{-1}_{T\rightarrow 0}/(T_1T)^{-1}_{T=T_c}}$. 
This relationship enables us to deduce $N_{\rm res}/N_0\sim$0.33 for $x$=1 and $N_{\rm res}/N_0\sim$0 for $x$=0. 
The solid line for $x$=1 in Fig.~\ref{1/T1TvsT}(b) is a tentative simulation based on the multiple-gap $s_\pm$-wave model. In this model, the nodal gap is on one of the multiple bands that is responsible for the RDOS and a larger gap with 2$\Delta_L/k_BT_c$=3.7 is on other bands that are mainly responsible for SC~\cite{Dulguun}. 

\begin{figure}[thp]
\centering
\includegraphics[width=8cm]{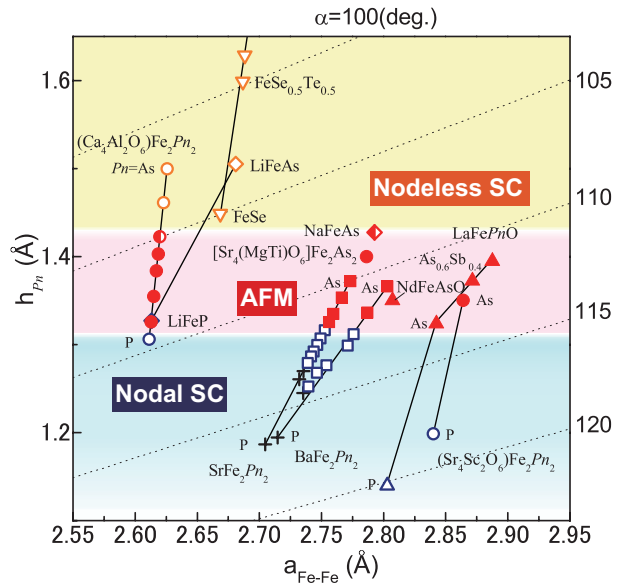}
\caption[]{(Color online) 
Map of the AFM ordered phase (filled symbols) and the SC phase (empty symbols) for (Ca$_4$Al$_2$O$_{6}$)Fe$_2$(As$_{1-x}$P$_x$)$_2$ plotted in the two-dimensional plane of structural parameters of the Fe-Fe distance $a_{\rm Fe-Fe}$ and $h_{Pn}$. Here, the emergent phases for various Fe-based compounds with the Fe$^{2+}$ state in (Fe$Pn$)$^-$ layer through the isovalent substitution at pnictogen ($Pn$) sites are presented with respect to Fe(Se,Te)\cite{Hanaguri,Kotegawa}, LiFe(As,P)\cite{Kim,Hashimoto}, BaFe$_2$(As,P)$_2$\cite{Kasahara,Huang,NakaiPRB,Yamashita,Zhang}, LaFe(Sb,As,P)O\cite{Cruz,Carlsson,Fletcher,Hicks}, NdFeAsO\cite{Qui}, and (Sr$_4M_2$O$_6$)Fe$_2$(As,P)$_2$ ($M$=Mg$_{0.5}$Ti$_{0.5}$\cite{Sato,Yamamoto},Sc\cite{Munevar,Yates}). The symbol ($+$) denotes the compounds which are not superconductive. The AFM order taking place universally in the range of $1.32$~\AA $\leq h_{Pn}\leq 1.42$\AA \ intervenes between the nodeless SC in $h_{Pn}>1.42$\AA \ and the nodal SC in $h_{Pn}<1.32$\AA . Dotted line is a linear relation of $h_{Pn}$ versus $a_{\rm Fe-Fe}$ at each value of $\alpha $.}
\label{AFM-SC_map}
\end{figure}

The present NMR studies on Al-42622(As,P) have revealed that as the P-substitution $x$ increases, the nodeless SC state with $T_c$= 27$\sim$20 K in 0$\le x \le$0.4 evolves into the AFM state in 0.5$\le x \le$0.95, and eventually to the nodal SC state with $T_c$=17 K at $x$=1 as presented in Fig.~\ref{Phase}. In another context, the AFM order intervenes between the nodeless SC and the nodal SC. This phase diagram is quite outstanding among numerous Fe-pnictides reported thus far. For example, it was reported that only the nodal SC state emerges in Ba122(As,P)~\cite{NakaiPRB,Yamashita,Zhang}.
To shed light on the occurrence of AFM order in the intermediate P-substitution range of 0.5$\le x \le$0.95, the AFM ordered state and the SC one for Al-42622(As,P) are plotted in the two-dimensional plane of structural parameters of the nearest-neighbour Fe-Fe distance $a_{\rm Fe-Fe}$ and $h_{Pn}$ by filled and empty symbols in Fig.~\ref{AFM-SC_map}, respectively. Here, the results are also presented on this plane with respect to 1111\cite{Cruz,Carlsson,Qui}, 122\cite{Kasahara,Huang}, 111\cite{Chen,Kitagawa}, and 42622-based compounds\cite{Sato,Yamamoto,Munevar} with the Fe$^{2+}$ state in the (Fe$Pn$)$^-$ layer through the isovalent substitution at pnictogen sites.
As a result, it is demonstrated that the AFM order in the (Fe$Pn$)$^-$ layer takes place universally irrespective of materials  when $h_{Pn}$ is in the range of $1.32$\AA $\leq h_{Pn}\leq 1.42$\AA . 
It is noteworthy that when $h_{Pn}>1.42$\AA , the nodeless SC emerges in Fe(Se,Te)~\cite{Hanaguri,Kotegawa} and LiFeAs\cite{Kim,Hashimoto} with the Fe$^{2+}$ state in the (Fe$Pn$)$^-$ layer as well as the case for Al-42622(As,P) with  0$\le x \le$0.4. On the other hand, when $h_{Pn}<1.32$\AA , the nodal SC takes place for $Ae$122(As,P) ($Ae$=Ba,Sr)\cite{Kasahara,NakaiPRB,Yamashita,Zhang,Dulguun}, LiFeP\cite{Hashimoto}, LaFePO\cite{Fletcher,Hicks}, and (Sr$_4$Sc$_2$O$_6$)Fe$_2$P$_2$\cite{Yates}. 
In this context, the quite unique and important ingredient found in Al-42622(As,P) is  that the nodeless SC and the nodal SC are separated by the onset of  AFM order. According to this empirical rule, it is reasonably understood that in the parent (111) compounds without any chemical doping, LiFeAs, and LiFeP~\cite{Kim,Hashimoto} exhibit the nodeless and nodal SC, respectively, whereas NaFeAs exhibits the AFM order~\cite{Chen,Kitagawa}. 
The novel two-dimensional map of the AFM ordered phase and the SC phase of Fig.~\ref{AFM-SC_map} is universal irrespective of a material's details, pointing to the importance of correlation effect for the emergence of SC as well as AFM order.


Finally, the present results on Al-42622(As,P) are considered in terms of a systematic evolution of Fermi-surface (FS) topology as the function of $h_{Pn}$ according to the band calculations based on the five-orbital model reported previously~\cite{Miyake,Usui,Kuroki2,Kosugi}. In general, the Fe-pnictides have similar FSs composed of disconnected two-dimensional hole pockets around $\Gamma$(0,0) and $\Gamma^{\prime}$($\pi$,$\pi$), and electron pockets around $M$[(0,$\pi$) and ($\pi$,0)] points. For the case of $x$=0 with a very large $h_{Pn}$=1.5 \AA~due to a narrow As-Fe-As bond angle $\alpha\sim~102^{\circ}$, a hole FS around $\Gamma^{\prime}$  is quite visible at the Fermi level, whereas one of the two-hole FSs at $\Gamma$ is missing~\cite{Miyake}, and hence the FS nesting condition is much better than in others~\cite{Miyake,Usui}. 
By contrast, for the case of $x$=1 with a very small $h_{Pn}$=1.31~\AA~due to $\alpha\sim~109.5^{\circ}$, the two-hole FSs at $\Gamma$ and one hole FS at $\Gamma^{\prime}$ appear  as well as  in $Ln$1111 with $T_c>$50 K~\cite{Kuroki2,Kosugi}. 
Although the FS multiplicity for x=1 is larger than that for x=0, the nesting condition of FSs in x=1 is worse than in x=0, bringing about the reduction in the Stoner factor for AFM correlations more significantly~\cite{Usui} as confirmed from the result in Fig.~\ref{1/T1TvsT}(a). 
In the intermediate region of $x$, the band calculation suggests the slight development of $\chi_Q$ and 
the reduction of eigenvalue in Eliashberg equation for s$_{\pm}$-wave pairing around $x\sim$0.7 \cite
{Kosugi,Kuroki_privatecom}. 
As for the nodal SC at $x$=1, although its $h_{Pn}$ is comparable to those of $Ae$122(As,P)($Ae$=Ba,Sr)~\cite{Kasahara,Dulguun}, the origin of the nodal SC  may differ from that in 122 compounds because of a highly two-dimensional electronic structure in the (Fe$Pn$)$^-$ layer separated by a thick perovskite-type block for Al-42622P. 

In conclusion, the $^{31}$P and $^{75}$As-NMR studies on (Ca$_4$Al$_2$O$_{6}$)Fe$_2$(As$_{1-x}$P$_x$)$_2$  have revealed the novel phase diagram including the nodeless SC ($0 \le x \le 0.4$) and the nodal SC ($x$=1) intimately separated by the onset of commensurate AFM order ($0.5 \le x \le 0.95$). It is highlighted that as a result of the fact that the P-substitution for As decreases the pnictogen height from the Fe plane,  the AFM order taking place in the range of $1.32$~\AA $\leq h_{Pn}\leq 1.42$~\AA  intervenes between the nodeless SC and the nodal SC and  this event is universal irrespective of materials with the Fe$^{2+}$ state in the (Fe$Pn$)$^-$ layer. 
In this context, the $s_\pm$-wave SC scenario mediated by spin fluctuations is quite promising when noting that this model 
has consistently accounted for our systematic experiments on series of compounds such as 42622, 1111, 122, and others reported thus far.  


{\footnotesize 
We thank K. Kuroki for valuable discussion and comments. This work was supported by a Grant-in-Aid for Specially Promoted Research (20001004) and by the Global COE Program (Core Research and Engineering of Advanced Materials-Interdisciplinary Education Center for Materials Science) from the Ministry of Education, Culture, Sports, Science and Technology (MEXT), Japan.
}


\end{document}